\documentclass[useAMS,usenatbib,usegraphicx]{mn2e}
\usepackage{color}
\usepackage{graphicx}
\usepackage{amssymb}

\def \apj {ApJ}
\def \apjl {ApJL}
\def \apjs {ApJS}

\def \mnras {MNRAS}
\def \pasj {PASJ}
\def \pasp {PASP}

\def \aap {A\&A}
\def \araa {ARA\&A}

\title[Low luminosity X-ray pulsators]{Near-infrared/optical identification of five low-luminosity X-ray pulsators$\thanks{Based on observations 
obtained from ESO Science Archive Facility under programs 66.D-0440(B) and 69.D-0339(A)}$
}
\author[Kaur et al.]{Ramanpreet Kaur$^{1}$\thanks{E-mail: r.kaur@uva.nl}, 
Rudy Wijnands$^{1}$, 
Biswajit Paul$^2$,  
Alessandro Patruno$^{1}$,
\newauthor 
Nathalie Degenaar$^{1}$\\
$^1$ Astronomical Institute "Anton Pannekoek", University of Amsterdam, 
Science Park 904, 1098 XH, Amsterdam, The Netherlands \\
$^2$ Raman Research Institute, C. V. Raman Avenue, Sadashivanagar, Bangalore 560 080, India.\\}
\begin{document}

\pubyear{2009}

\maketitle

\label{firstpage}

\begin{abstract}

We present the identification of the most likely near-infrared/optical 
counterparts of five low-luminosity X-ray pulsators (AX J1700.1--4157, AX J1740.1--2847, 
AX J1749.2--2725, AX J1820.5--1434 and AX J1832.3-0840) which have long 
pulse periods ($>$ 150 s). The X-ray properties of these systems suggest that they
are likely members of persistent high mass X-ray binaries or intermediate polars.
Using our \emph{Chandra} observations, we detected the most likely 
counterparts of three sources (excluding AX J1820.5--1434 and AX J1832.3-0840) 
in their ESO - NTT near-infrared observations, 
and a possible counterpart for AX J1820.5--1434 and AX J1832.3--0840 in the 2MASS 
and DSS observations respectively. 
We also performed the X-ray timing and spectral analysis for all the 
sources using our \emph{XMM-Newton} observations, which further helped 
us to constrain the nature of these systems. Our multiwavelength
observations suggest that AX J1749.2 -2725 and AX J1820.5 -1434 most
likely harbor accreting neutron stars while AX J1700.1 -4157, AX J1740.1--2847 and 
AX J1832.3--0840 could be intermediate polars.

\end{abstract}


\section{Introduction}

During the last decade, a number of Galactic sub-luminous X-ray pulsators 
(L$_X$ $\sim$ 10$^{34}$  - 10$^{35}$ erg s$^{-1}$) have been discovered  
with pulse periods ranging from a few seconds to over a thousand
seconds. Most of these pulsators are discovered during Galactic plane
surveys performed using various X-ray telescopes, e.g., \emph{BeppoSAX}, 
\emph{ASCA}, and \emph{RXTE}, 
and are found to harbor a variety of source-types like
anomalous X-ray pulsars (isolated slowly rotating neutron stars; e.g., 
\citealt{Torii1998}), accreting magnetized white dwarfs (i.e., intermediate 
polars (IPs); e.g., \citealt{Misaki1996}) and neutron stars accreting from
high mass companions (i.e., mostly Be/X-ray binaries; e.g., \citealt{Hulleman1998}).
A majority of these systems are transient in nature, but there
are several persistent systems also emitting at such low-luminosities. Among these 
persistent systems, there remains a group of sources whose nature  has not been
determined yet. The X-ray properties of most of these persistent systems suggest  
that they could be neutron stars accreting from a high mass star or accreting white 
dwarf systems.
Furthermore, in some cases, the possibility of a system in which a neutron
star accretes from a low mass star also cannot be excluded \citep{Lin2002}. 

Most of the known Be/X-ray binaries are transient in nature and this
behaviour is usually associated with their highly eccentric orbits 
\citep{Okazaki2001}. A sub-class of Be/X-ray binaries characterized by persistent, 
low-luminosity X-ray emission, and slowly rotating pulsars, has recently been
proposed by \citet{Reig1999}.
The neutron star in these systems is assumed to 
be orbiting its companion Be star in a relatively wide and a circular orbit, 
hence accreting from the low-density outer regions of the circumstellar envelope. 
\citet{Pfahl2002} has proposed a possible scenario for the formation of these 
wide orbit ($>$ 30 days) and low-eccentricity 
($<$ 0.2) high mass X-ray binaries, suggesting 
that these systems could have formed in a supernova explosion, accompanied with 
a very small kick ($\lesssim$ 50 km/s) to the neutron star. If true, this growing
class of persistent Be/X-ray binaries would help us to explore a different type 
of supernova explosion.

This paper is a part of our ongoing project to find the 
true nature of low-luminosity X-ray pulsators. 
In our previous paper, we reported the identification 
of NIR counterparts of two low-luminosity X-ray pulsators
- SAX J1324.4-6200 and SAX J1452.8-4959 using the \emph{Chandra} and 
the \emph{XMM-Newton} observations \citep{Kaur2009}. It was
suggested that SAX J1324.4-6200 is likely a HMXB pulsar 
while no firm conclusion about the nature of SAX J1452.8-4959 could be drawn.
In this paper, we report on the identification of the most likely 
counterparts of additional five low-luminosity X-ray pulsators - AX J1700.1-4157, 
AX J1740.1-2847, AX J1749.2-2725, AX J1820.5-1434 and AX J1832.3-0840. 
These sources were discovered from the Galactic plane observations 
made in 1995 - 1999 using the \emph{ASCA} satellite \citep{Sugizaki2001}. 
The basic physical parameters with which these sources were discovered are listed 
in Table \ref{tt1}.

\begin{table*}
\centering
\caption{Summary of basic parameters of our sample of X-ray pulsators.}
\medskip
\small
\label{tt1}
\begin{tabular}{lllllccc}
\hline
\hline
Object    &  $l$ & $b$ & Spin period           & Absorbed flux & \multicolumn{2}{c}{Spectral parameters} & References\\
        &    &   & (in seconds)     & ($\times$ 10$^{-12}$ erg cm$^{-2}$ s$^{-1}$)  &  N$_\mathrm{H}$ & \\
                  & && & (2 - 10 keV) & ($\times$ 10$^{22}$ cm$^{-2}$) & $\Gamma$ & \\ \hline
AX J1700.1--4157  & 344.04 &  0.24  & 714.5  $\pm$ 0.3     & 6        &  6  & 0.7 &   1 \\
AX J1740.1--2847  & 359.49 &  1.09  & 729    $\pm$ 14      & 4        &  3  & 0.7 &   2 \\
AX J1749.2--2725  & 1.70 & 0.11     & 220.38 $\pm$ 0.20    & 3 - 30   &  10 & 1.0 &   3 \\
AX J1820.5--1434  & 16.47 & 0.07    & 152.26 $\pm$ 0.04    & 23       &  10 & 0.9 &   4 \\
AX J1832.3--0840  & 23.04 & 0.26    & 1549.1 $\pm$ 0.4     & 11       &  1  & 0.8 &   5 \\ \hline
\end{tabular} 
\flushleft
REFERENCES. -- (1) \citealt{Torii1999} (2) \citealt{Sakano2000} (3) \citealt{Torii1998} 
(4) \citealt{Kinugasa1998} (5) \citealt{Sugizaki2000} \\
NOTE : In this table, $l$  and $b$ represent the Galactic longitude and latitude of a given star respectively.
The spectral parameters, N$_\mathrm{H}$ and $\Gamma$, represent the neutral hydrogen column density 
and the power-law index respectively. 

\end{table*}

\section{X-ray observations}

We carried out X-ray observations of the five X-ray pulsators listed in Table \ref{tt1} 
using the European Photon Imaging Camera (EPIC) aboard the \emph{XMM-Newton} satellite 
and the Advanced CCD Imaging Spectrometer (ACIS) aboard the \emph{Chandra} satellite. 

\subsection{\it{XMM-Newton}}

The X-ray pulsators were observed for 7 - 32 ks
each using the EPIC instruments on the \emph{XMM-Newton} satellite.
The observation details are summarized in Table \ref{t1}.
During our observations, both the EPIC-\emph{MOS} and \emph{pn}
cameras (\citealt{Turner2001}; \citealt{Struder2001}) were operated in the
\emph{Full Frame} mode and with the medium filter. The EPIC 
data was processed using the \emph{XMM-Newton} Science
Analysis System {(SAS version 7.1.0)\footnote[1]
{see http://xmm2.esac.esa.int/sas/}}.

Investigation of the full-field count-rate of X-ray pulsators
revealed no high-count rate background particle flaring
in all of them so all data could be used. During our observations
for all the targets, we detected only one source inside the error 
circle from the \emph{ASCA} observations.
The detection of pulsations in all of them (see Section \ref{s6}) 
further confirmed their identity.
We used the task 
\emph{edetect\_chain} to find the exact position of the X-ray sources  
(see Table \ref{t1})
in their combined EPIC-\emph{MOS} and \emph{pn} image. 
The error circle on the position of each X-ray source
is adopted as a quadratic sum of the bore
sight error of the \emph{XMM-Newton} {telescope\footnote[2]
{see http://xmm2.esac.esa.int/docs/documents/CAL-TN-0018.pdf}
and the statistical error given by the task \emph{edetect\_chain} and is 
$\sim$ 2$\farcs$0 for all the targets.

\begin{table*}
\centering
\caption{X-ray observation details.}
\small
\medskip
\label{t1}
\begin{tabular}{llccccccc}
\hline
\hline
Object & Telescope &  Date & ObsId & Duration &    R.A.     & DEC.      \\
                     &        &  (UT) &       &  (ks)    &    hh:mm:ss & $^{\circ}$ $^{\prime}$ ${\arcsec}$ \\ \hline

AX J1700.1--4157   & \emph{XMM-Newton} & 17 Feb 2008 & 0511010601 &  7.9 &  17:00:04.32  &  $-$41:58:04.44    \\
                   & \emph{Chandra} & 30 Jun 2008 & 9015          & 1.13 &  17:00:04.35  &  $-$41:58:05.46    \\
AX J1740.1--2847   & \emph{XMM-Newton} & 27 Feb 2008 & 0511010701 &  9.3 &  17:40:09.12  &  $-$28:47:26.16    \\
                   & \emph{Chandra} & 06 Sep 2008 & 9016          & 1.17 &  17:40:09.12  &  $-$28:47:26.02    \\
AX J1749.2--2725   & \emph{XMM-Newton} & 04 Mar 2008 & 0511010301 &  8.9 &  17:49:12.24  &  $-$27:25:37.56    \\
                   & \emph{Chandra} & 27 Apr 2008 & 9013          & 1.18 &  17:49:12.41  &  $-$27:25:38.21    \\
AX J1820.5--1434   & \emph{XMM-Newton} & 30 Sep 2007 & 0511010101 & 11.2 &  18:20:30.00  &  $-$14:34:23.16    \\
                   & \emph{Chandra} & 03 Sep 2008 & 9011          & 1.15 &  18:20:30.09  &  $-$14:34:23.52    \\
AX J1832.3--0840   & \emph{XMM-Newton} & 16 Oct 2007 & 0511010801 & 31.3 &  18:32:19.44  &  $-$08:40:30.47    \\
                   & \emph{Chandra} & 07 Aug 2008 & 9017          & 1.16 &  18:32:19.30  &  $-$08:40:30.44    \\

\hline
\end{tabular}

NOTE : The error circle on the \emph{Chandra} and the \emph{XMM-Newton} observations is $\sim$ 
0$\farcs$64 and $\sim$ 2$\farcs$0 respectively for all the sources.\\
\end{table*}

\subsection{\it{Chandra}}

The X-ray pulsators were observed for $\sim$ 1 ks each using the ACIS-I instrument 
on the \emph{Chandra} satellite in the \emph{FAINT} mode.
The details of these observations are summarized in Table \ref{t1}.
We processed the ACIS-I event 2 files using the standard
software packages CIAO {4.0\footnote[3] {\emph{Chandra} Interactive Analysis
of Observations (CIAO), http://cxc.harvard.edu/ciao/.}} and CALDB
{3.4.2\footnote[4] {\emph{Chandra} Calibration Database (CALDB),
http://cxc.harvard.edu/caldb/}. The task \emph{wavdetect} was used to
find the exact position of the targets in the ACIS-I images and
are listed in Table \ref{t1}. The error circle on the position 
is adopted as a quadratic sum of the bore sight error of the \emph{Chandra}
{telescope\footnote[5] {see http://cxc.harvard.edu/cal/ASPECT/celmon.} (0$\farcs$6, \citealt{Aldcroft2000}),
1-$\sigma$ {\it wavdetect} errors and a contribution that depends on
the number of detected counts \citep{vandenberg2004} and is found to 
be $\sim$ 0$\farcs$64 for all the sources.

\section{X-ray timing analysis}
\label{s6}

The \emph{XMM-Newton} EPIC-\emph{MOS} and \emph{pn} observations are used for the
timing analysis of our targets. The X-ray events
were extracted in a circular region of radius 30$\arcsec$ centered on the
position of the X-ray source in their EPIC-\emph{MOS} and \emph{pn} images,
with a time resolution of 0.1 s. Similarly, the background X-ray events 
were extracted for each source from a source-free region on the 
same CCD. The event times were then transformed to barycentric times
using the \emph{Chandra} position of the X-ray source and the
JPL-DE405 ephemeris using the task \emph{barycor} in \emph{SAS}.
The background was subtracted to generate the background 
corrected light curves separately for EPIC-\emph{MOS} and 
\emph{pn} instruments, which were then added to generate 
the final lightcurves for our timing analysis.

We searched for the spin-period in the lightcurves using the task \emph{powspec} in 
\emph{FTOOLS}\footnote[6]{http://heasarc.gsfc.nasa.gov/docs/software/ftools/ftools\_menu.html}
and refined it further using the task \emph{efsearch}. 
The pulse periods thus detected for all pulsators are listed in Table \ref{t2}
except for AX J1820.5--1434. We phase connected the pulsations
for AX J1820.5--1434, using EPIC - \emph{pn} observations,  
by cross-correlating each single profile with a standard profile 
obtained by folding the entire data set. 
This method gave us a pulse period of AX J1820.5--1434 with a much better 
accuracy, P$_\mathrm{s}$ = 153.24 $\pm$ 0.02 s (listed in Table \ref{t2}).   
The errors on the pulse periods are calculated at the 68\% confidence level. 
Given the short-baseline of all the observations, it was not possible 
to measure the spin period derivative ($\dot{P}$) for all the 
sources with a phase coherent 
technique. However, we fitted the previous (see Table \ref{tt1}) 
and the new measured values of the spin-period (see Table \ref{t2})
with a linear relation and determined the spin-period derivative of 
AX J1820.5--1434 to be (3.00 $\pm$ 0.14) $\times$ 10$^{-9}$ s s$^{-1}$ 
and an upper limit for the remaining sources (see Table \ref{t2}). 
Assuming an orbital velocity of the pulsating component to be $\sim$ 300 km s$^{-1}$, 
the difference in pulse period ($\Delta{P_\mathrm{s}}$) measured in different orbital 
phases would be $\sim$ P$_\mathrm{s}$/1000. However, in case of  AX J1820.5--1434,
the measured $\Delta{P_\mathrm{s}}$ is 0.98 s ($\sim$ P$_\mathrm{s}$/100),  
indicates that the measured period derivative is significantly more than that 
caused by the orbital motion.

\begin{figure*}
\centering
\medskip
\includegraphics[height= 17.0cm,width= 14cm]{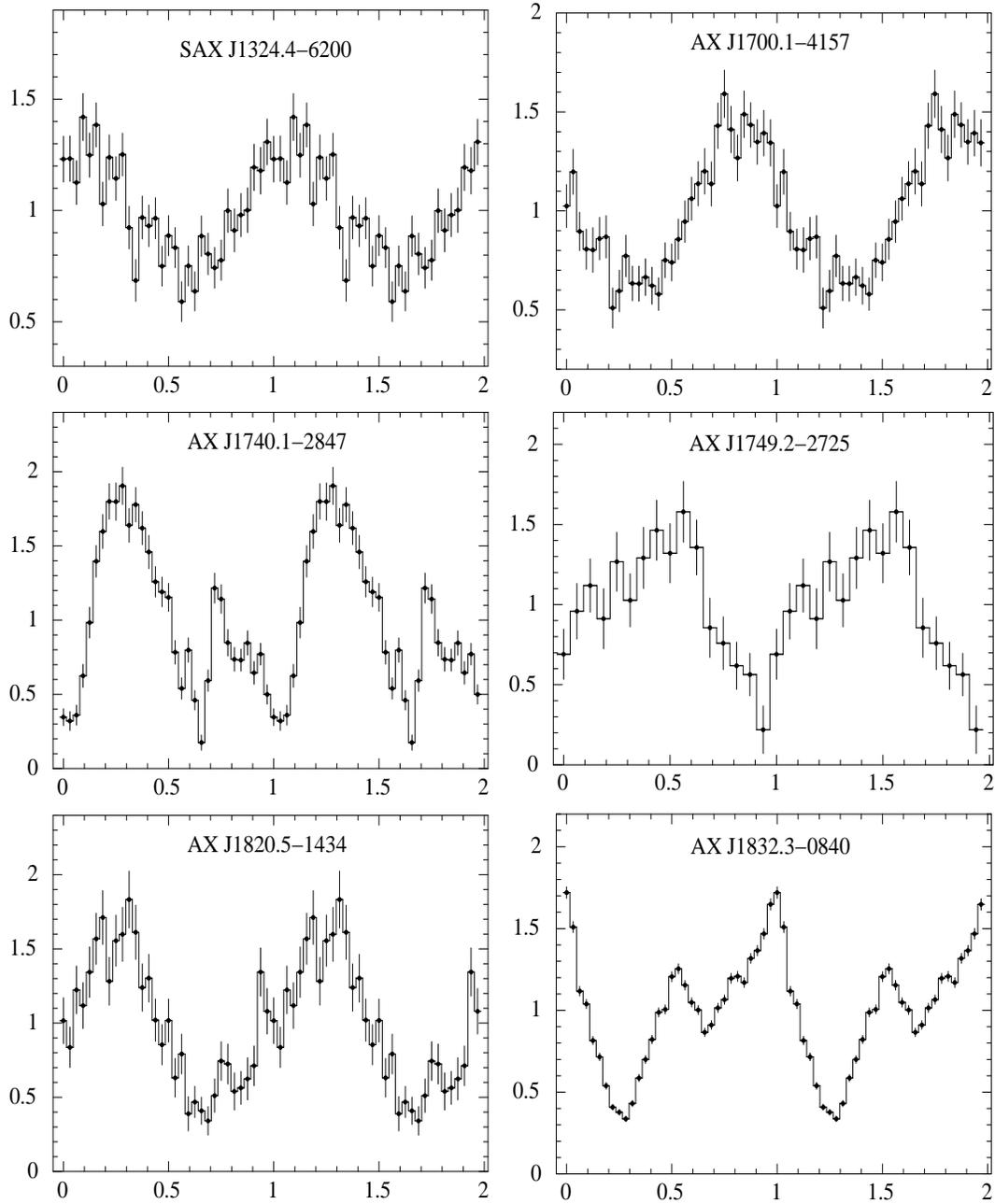}
\caption{Pulse profiles of the X-ray pulsators from   
their \emph{XMM-Newton} observations. The pulse 
profile of SAX J1324.4--6200 from \citet{Kaur2009} is also included 
for comparison.} 
\label{f1}
\end{figure*}

The obtained X-ray lightcurves are folded in one single profile to measure
the pulse fractional amplitude defined as  
(F$_{max}$ - F$_{min}$)/(F$_{max}$ + F$_{min}$)
where F$_{max}$ and F$_{min}$ are the maximum and minimum flux (or counts)
in a pulse profile. The pulse profiles of all the sources thus obtained are
shown together in Figure \ref{f1} where the pulse profile of
SAX J1324.4-6200 is taken from \citet{Kaur2009} for comparison. 
Out of the six sources listed in Table \ref{t2}, four sources clearly 
showed single peak profiles, 
while AX J1740.1-2847 and AX J1832.3--0840 showed double peak profiles.

\begin{table*}
\centering
\caption{The pulse period, pulse fractional amplitude (\%) and the 
spin period derivative, $\dot{\mathrm{P}}$ of our targets. 
The upper limits are quoted with 95\% confidence level. 
SAX J1324.4--6200  is included from \citet{Kaur2009} for a comparison.} 
\medskip
\label{t2}
\begin{tabular}{lllc}
\hline
\hline
Object            &  Pulse period             & Pulse fractional &  $\dot{\mathrm{P}}$  \\
                  &  (seconds)                &  amplitude (\%)  & ($\times$10$^{-9}$ s s$^{-1}$)\\
\hline
SAX J1324.4--6200 &  172.57 $\pm$ 0.22 &   52 $\pm$ 4             & 6.34 $\pm$ 0.08 \\
AX J1700.1--4157  &  723.7$^{+13.1}_{-1.7}$   &   52 $\pm$ 8             & $<$ 84 \\
AX J1740.1--2847  &  730.5$^{+5.6}_{-1.0}$    &   81 $\pm$ 12            & $<$ 93 \\
AX J1749.2--2725  &  218.1$^{+1.3}_{-1.9}$    &   77 $\pm$ 13            & $<$ 1  \\
AX J1820.5--1434  &  153.24 $\pm$ 0.02        &   65 $\pm$ 11            & 3.00 $\pm$ 0.14 \\
AX J1832.3--0840  &  1552.3$^{+2.3}_{-0.8}$   &   66 $\pm$ 3             & $<$ 18 \\
\hline
\end{tabular}
\end{table*}

\section{X-ray spectral analysis}
\label{s7}

Using the \emph{XMM-Newton} data, we also performed spectral analysis of our targets.
We used the same extraction region for both these instruments 
as were used for the timing analysis reported in Section \ref{s6}.
The SAS tool {\it xmmselect} was used to extract both the source and the background
spectra and {\it XSPEC} (version 12.0.0) was used for the spectral fitting.
The resulting spectra were re-binned to have minimum of 20 counts per bin.

\begin{figure*}
\centering
\medskip
\includegraphics[height=19.0cm,width=17cm]{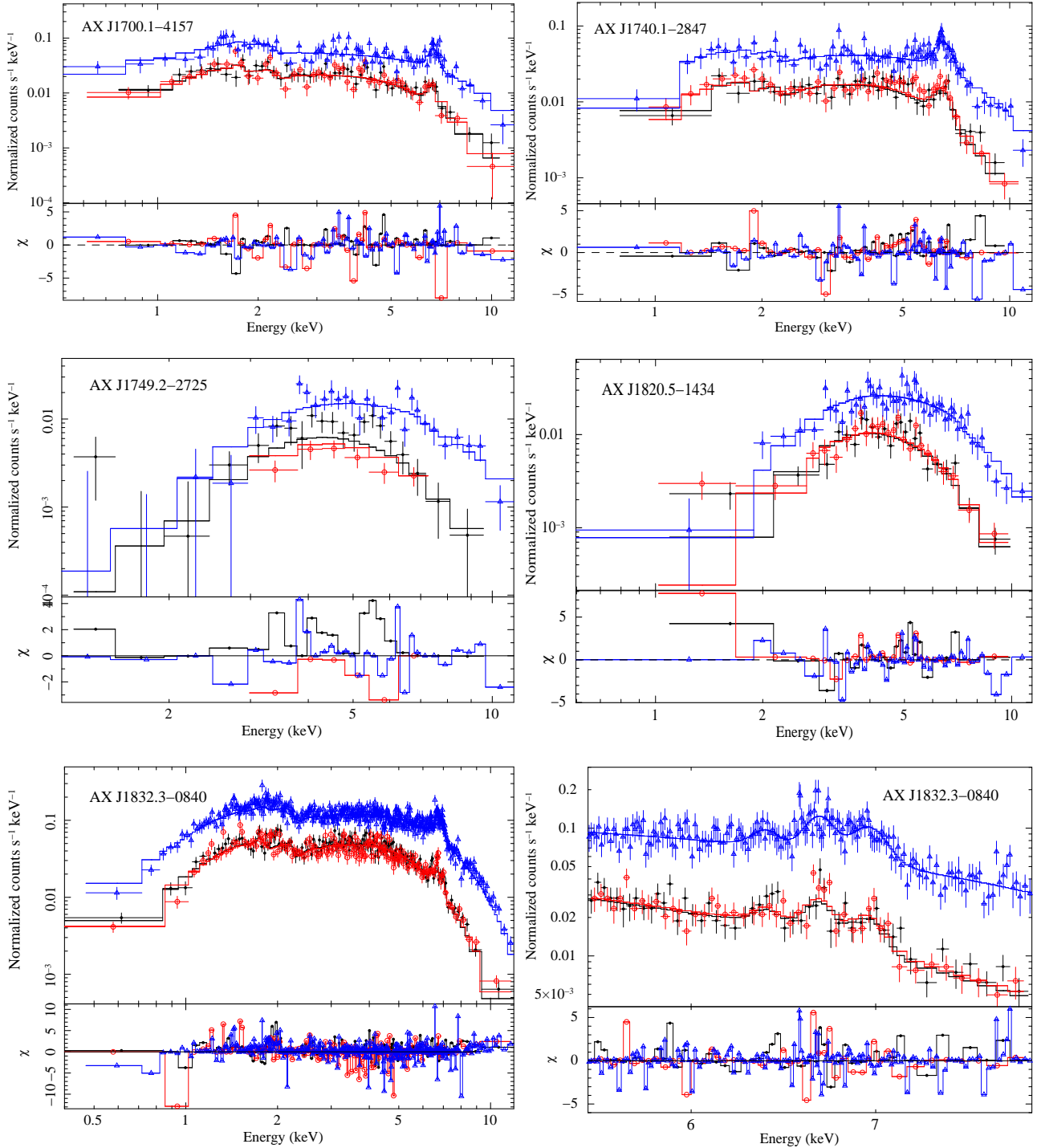}
\caption{The \emph{XMM-Newton} EPIC-\emph{MOS} and \emph{pn} spectra of X-ray pulsators 
fitted with an absorbed powerlaw model except for AX J1832.3--0840, where an
absorbed blackbody model 
fit is shown. The EPIC-\emph{pn}, \emph{MOS1} and \emph{MOS2} data points are represented 
by open triangles (blue color), open circles (red color) and filled circles (black color) 
respectively. A closer view of the X-ray spectrum of AX J1832.3--0840 from 5 - 8 keV is 
also provided to have a better look at different Fe emission lines.}
\label{f2}
\end{figure*}

We fitted a variety of single component models to the spectra. 
The X-ray spectra of most of our sources could be fitted well using 
the absorbed power-law and the blackbody models.
For all pulsators, the values of the measured fit parameters using these two models 
are given in Table \ref{t3}.
The absorbed as well as the unabsorbed fluxes for the fitted models in
2 - 10 keV energy band are also given in the same table.
Figure \ref{f2} shows the best fit powerlaw model to the X-ray spectra 
of our targets, except for AX J1832.3-0840, where the blackbody model fit
is shown. For all the sources, the thin thermal plasma model 
(\emph{MEKAL}; \citealt{Mewe1995}) gave a poor fit while the bremsstrahlung 
model with a partial absorption component gave an acceptable 
fit only for AX J1832.3--0840 with a reduced $\chi^2$ of 1.1 for 1250 degrees of
freedom.  The parameters obtained from this fit are as follows :
bremsstrahlung temperature,
kT = 29.4$^{+10.6}_{-6.9}$, interstellar absorption, N$_\mathrm{H}$ =
0.81 $\pm$ $0.05$ cm$^{-2}$, partial absorption, N$_\mathrm{H1}$ =
6.2$^{+0.9}_{-0.8}$ cm$^{-2}$ and the partial covering fraction = 0.68 $\pm$ $0.02$.
The detection of an Fe emission lines is marked as `y' in
Table \ref{t3} and the further details are given
in Table \ref{t4}. In the case when no Fe emission line is detected, 
an upper limit on Fe 6.4 keV line flux is given in Table \ref{t4}.
This was calculated by fixing the line-center at 6.4 keV and 
line-width at 0.1 keV in the spectrum and fitting a Gaussian to the line. The 
parameters of Fe lines reported in Table \ref{t4} are measured using 
the power-law model fit to the spectra except for AX J1832.3--0840, for which 
blackbody model fit is used. 
Using the observed X-ray flux, we have also estimated the X-ray luminosity 
of our targets at a distance of 1 kpc (typical for IPs) and 8 kpc 
(typical for X-ray binaries) and these are listed in Table \ref{t3}.

\begin{table*}
\centering
\caption{The X-ray spectral parameters of our pulsators, measured using the \emph{XMM-Newton} 
observations.}
\medskip
\label{t3}
{\scriptsize
\begin{tabular}{lccccccc}
\hline \hline
&   AX J1700.1--4157 & AX J1740.1--2847 & AX J1749.2--2725 & AX J1820.5--1434 & AX J1832.3-0840 \\  \hline
Power-law  &&&&&&& \\ \hline
N$_\mathrm{H}$ ($\times$ 10$^{22}$ cm$^{-2}$)  & 0.52$^{+0.16}_{-0.12}$ & 1.0 $\pm$ 0.2 & 13.2$^{+4.1}_{-3.5}$& 8.4$^{+2.8}_{-1.0}$&  0.99 $\pm$ 0.05 \\
$\Gamma$ &  0.56$^{+0.13}_{-0.11}$ & 0.5 $\pm$ 0.1 & 1.5$^{+0.6}_{-0.5}$ &1.41$^{+0.43}_{-0.16}$ & 0.83 $\pm$ 0.03 \\
$F_{X, \mathrm{abs}}$ ($\times$ 10$^{-12}$ erg s$^{-1}$ cm$^{-2}$) & 3.7 $\pm$ 1.2  & 3.4 $\pm$ 0.8 & 1.1 $\pm$ 1.7 & 1.6 $\pm$ 1.2 & 7.1 $\pm$ 0.5 \\
$F_{X, \mathrm{unabs}}$ ($\times$ 10$^{-12}$ erg s$^{-1}$ cm$^{-2}$) & 3.8 $\pm$ 1.2 & 3.5 $\pm$ 0.8 & 2.2 $\pm$ 3.0 & 2.5 $\pm$ 0.7 & 7.5 $\pm$ 0.5 \\
Fe line & y & y & - & - & y \\
$\chi_{\nu}^2/\nu$ & 1.1/245 & 1.0/205 & 1.1/55 & 1.0/115 & 1.4/1250  \\
L$_X$ at 1 kpc ($\times$10$^{32}$ erg s$^{-1}$) & 4.4 $\pm$ 1.4 & 4.0 $\pm$ 1.0 & 1.3 $\pm$ 2.0 & 1.9 $\pm$ 1.4 & 8.5 $\pm$ 0.6 \\
L$_X$ at 8 kpc ($\times$10$^{34}$ erg s$^{-1}$) & 2.8 $\pm$ 0.9 & 2.6 $\pm$ 0.6 & 0.9 $\pm$ 1.3 & 1.2 $\pm$ 0.9 & 5.4 $\pm$ 0.4 \\ \hline
Black-body   &&&&&&&    \\ \hline
N$_\mathrm{H}$ ($\times$10$^{22}$ cm$^{-2}$) & $<$ 0.064 & 0.2 $\pm$ 0.1 & 7.6$^{+3.0}_{-2.3}$ & 4.6$^{+1.1}_{-0.9}$ &  0.23 $\pm$ 0.02   \\
$kT$ (keV) & 1.93$^{+0.12}_{-0.12}$ & 2.4 $\pm$ 0.1 & 2.1$^{+0.4}_{-0.3}$ & 1.92$^{+0.17}_{-0.15}$& 1.90 $\pm$ 0.03 \\
$F_{X, \mathrm{abs}}$ ($\times$ 10$^{-12}$ erg s$^{-1}$ cm$^{-2}$) & 3.3 $\pm$ 0.2 & 3.2 $\pm$ 0.4 & 1.1 $\pm$ 0.2 & 1.6 $\pm$ 0.1 & 6.8 $\pm$ 0.1 \\
$F_{X, \mathrm{unabs}}$ ($\times$ 10$^{-12}$ erg s$^{-1}$ cm$^{-2}$)  & 3.3 $\pm$ 0.2 & 3.3 $\pm$ 0.4 & 1.5 $\pm$ 0.3 & 1.9 $\pm$ 0.1& 6.9 $\pm$  0.1     \\
Fe line &  y & y & - &- & y \\
$\chi_{\nu}^2/\nu$ & 1.0/245 & 0.9/205 & 1.2/55 & 0.9/115 & 1.1/1250         \\ 
L$_X$ at 1 kpc ($\times$10$^{32}$ erg s$^{-1}$) & 3.9 $\pm$ 0.2 & 3.7 $\pm$ 0.5 & 1.3 $\pm$ 0.2 & 1.9 $\pm$ 0.1 & 8.1 $\pm$ 0.1 \\
L$_X$ at 8 kpc ($\times$10$^{34}$ erg s$^{-1}$) & 2.5 $\pm$ 0.2 & 2.3 $\pm$ 0.3 & 0.8 $\pm$ 0.2 & 1.2 $\pm$ 0.1 & 5.2 $\pm$ 0.1 \\ \hline
\end{tabular}}
\flushleft
Note : For all the sources, the observed fluxes ($F_{X, \mathrm{abs}}$), unabsorbed 
fluxes ($F_{X, \mathrm{unabs}}$) and the X-ray luminosities (L$_X$) are measured in 
the energy band 2 - 10 keV. The sources in which we detected the Fe emission lines 
are marked as `y'. 
\end{table*}

\begin{table*}
\centering
\caption{Properties of the Fe emission lines detected in the X-ray spectrum of 
our targets using a power-law model except for AX J1832.3-0840, for 
which blackbody model fit is used. An upper limit 
on equivalent width (EW) of the Fe 6.4 keV line is given in case of non-detection 
of any Fe emission line in the X-ray spectrum.}
\medskip
\label{t4}
{\scriptsize
\begin{tabular}{lcccccccccc} \hline \hline
Object &   \multicolumn{3}{c}{Fe 6.40 keV (Flourescent)}  &   \multicolumn{3}{c}{Fe 6.67 keV (He-like)} &   \multicolumn{3}{c}{Fe 6.97 keV (H-like)} \\
    &        Center  & width  & EW &  Center & width & EW &  Center & width & EW \\  
    & (keV) & (keV) & (eV) & (keV) & (keV) & (eV) & (keV) & (keV) & (eV) \\ \hline
AX J1700.1--4157 & - & - & - & 6.69 $\pm$ 0.08  &  0.17 $\pm$ 0.09  &  580    &   -   &  -   &   -   \\
AX J1740.1--2847 & 6.5 $\pm$ 0.1 & 0.3 $\pm$ 0.1  &  998    &   -   &  -  &  -  &  -   &   -   & -  \\
AX J1749.2--2725 & 6.4 (frozen) & 0.1 (frozen) & $<$ 88 &-&-&-&-&-&-   \\
AX J1820.5--1434 & 6.4 (frozen) & 0.1 (frozen) & $<$ 82 &- & - & - & - & - & -  \\
AX J1832.3-0840  & 6.39 $\pm$ 0.03 & $<$ 0.07 & 50 & 6.67 $\pm$ 0.02  & $<$ 0.06 & 825 & 6.96 $\pm$ 0.03 & 0.06 $\pm$ 0.04 & 736  \\ \hline
\end{tabular}
}
\flushleft
\end{table*}

\section{Infrared observations and data analysis}

We retrieved near-infrared (NIR) observations of three sources - AX J1700.1--4157, AX J1740.1--2847
and AX J1749.2--2725 from the ESO - NTT archive and the details are 
summarized in Table \ref{t6}. 
These observations were made in NIR \emph{J}, \emph{H},
\emph{K$_{\mathrm{s}}$} filters with a NIR imager and spectrograph called
Son-of-ISAAC (\emph{SOFI}). 
The instrument was set up in large imaging mode with a pixel
scale of 0${\farcs}$29 and a FOV of 5$\arcmin$ $\times$ 5$\arcmin$
for AX J1700.1--4157 and AX J1740.1--2847, while in small imaging mode
with a pixel scale of 0${\farcs}$14 and a FOV of
2$^\prime$.5 $\times$ 2$^\prime$.5 for AX J1749.2--2725. 
During these observations, the seeing 
varied from $0{\farcs}6$ to $1{\farcs}0$. 
Images were acquired in the auto-jitter mode in which
a number of single frames (NDIT) having exposure
times of DIT (Detector Integrator Time) seconds were acquired
at different positions and then co-averaged to generate
an output image.

The data reduction is done using the standard routines 
in \emph{IRAF}\footnote[7]{IRAF is distributed by the 
National Optical Astronomy Observatories, USA.}.
A master sky-frame is first
constructed by median stacking all the frames of a source
for each filter, which is then subtracted from all the 
images to generate sky-subtracted frames. These images
are then flat-fielded, aligned and average stacked to
obtain the final images. The astrometry and calibration
of the final frames is performed using the 2MASS observations
which gave an (absolute) position uncertainty of $\sim$ 0${\farcs}$2 
for our observations.

\begin{table*}
\caption{Log of the near-infrared observations 
obtained using 3.52-m ESO - New Technology Telescope (NTT). NDIT represents the
number of single frames, having exposure times of DIT
(detector integrator time) seconds and are used to generate
an output image having exposure time equal to one DIT.
The number of output frames are represented by Nframes.}
\medskip
\tiny
\label{t6}
\centering
{\scriptsize
\begin{tabular}{lccccccccccc}
\hline
\hline
Source & Date & Program ID & \multicolumn{3}{c}{{\it J}} & \multicolumn{3}{c}{{\it H}} & \multicolumn{3}{c}{{\it K}$_{\mathrm{s}}$}  \\ \hline
& (UT)  & &  DIT & NDIT & Nframes & DIT & NDIT & Nframes & DIT & NDIT & Nframes \\
\hline
AX J1700.1--4157 & 20 Mar 2001 & 66.D-0440(B) & 3 & 5 & 4  & 3 & 5 & 4  & 3  & 5 & 6  \\
AX J1740.1--2847 & 18 Jun 2002 & 69.D-0339(A) & 3 & 20 & 5  & 3 & 20 & 5 & 3 & 20 &5  \\
AX J1749.2--2725 & 20 Mar 2001 & 66.D-0440(B) & 3 & 5 & 4 &  3 & 5 & 4 & 3 & 5 & 6 \\
 \hline
\end{tabular}}
\end{table*}

\begin{table*}
\centering
\caption{The observed \emph{J}, \emph{H} and \emph{K$_{\mathrm{s}}$} magnitudes of the most 
likely NIR counterparts of the X-ray pulsators. The R.A. and DEC. of these sources measured
from their NIR observations are also given. The position of these sources is measured
with an uncertainty of $\sim$ 0.2$\arcsec$.} 
\medskip
\label{t7}
\begin{tabular}{lccccc}
\hline
\hline
Star &  R.A.    & DEC.                                 & \emph{J}       & \emph{H} &  \emph{K$_{\mathrm{s}}$} \\
     &  hh:mm:ss&  $^{\circ}$ $^{\prime}$ ${\arcsec}$  & magnitude      & magnitude& magnitude \\\hline
AX J1700.1--4157  &17:00:04.35  & $-$41:58:05.52 &17.07 $\pm$ 0.04& 17.04 $\pm$ 0.05 & 17.38 $\pm$ 0.13\\
AX J1740.1--2847  &17:40:09.14  & $-$28:47:25.68 &16.16 $\pm$ 0.03& 15.76 $\pm$ 0.07 & 15.57 $\pm$ 0.10\\
AX J1749.2--2725  &17:49:12.41 & $-$27:25:38.25 & not detected   & 16.89 $\pm$ 0.07 & 15.15 $\pm$ 0.02\\
AX J1820.5--1434  &18:20:30.10  & $-$14:34:22.90  &15.41 $\pm$ 0.15& 13.25 $\pm$ 0.07 & 11.75 $\pm$ 0.04\\
AX J1832.3--0840$^{*}$  &18:32:19.39 & $-$08:40:30.53 &17.11 $\pm$ 0.23& 16.22 $\pm$ 0.21 & 16.06 $\pm$ 0.34\\
\hline
\end{tabular}

* - It is also detected in the Digitized Sky Survey (DSS) observations with magnitudes, R = 19.58 mag and B = 21.13 mag. \\
\end{table*}

\section{Identification of NIR/optical counterparts}

\begin{figure*}
\centering
\medskip
\includegraphics[height=22.0cm,width=10.0cm]{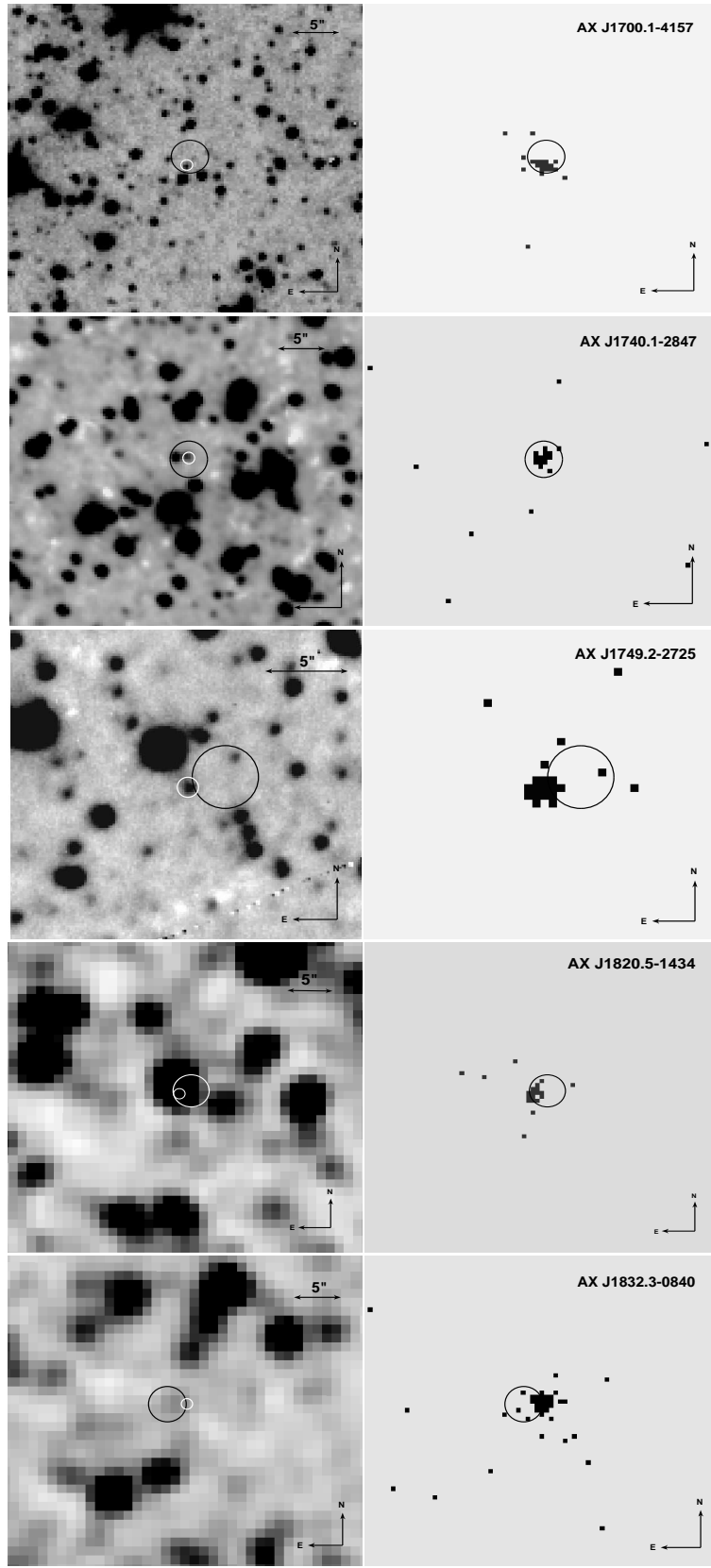}
\caption{\emph{Left} : The ESO-NTT NIR  \emph{K$_{\mathrm{s}}$}  waveband image of
AX J1700.1--4157, AX J1740.1--2847, AX J1749.2--2725 and 2MASS $K_\mathrm{s}$ 
waveband image 
of AX J1820.5--1434 and AX J1832.2--0840. \emph{Right} : The \emph{Chandra} 
ACIS-I images of the same sources. The black and the white circles 
represent error circles on their position obtained from their \emph{XMM-Newton} 
and the \emph{Chandra} observations respectively, except for the NIR 
observations of AX J1820.5--1434 for which larger white circle represents the 
error circle obtained from the \emph{XMM-Newton} observations.}
\label{f3}
\end{figure*}

We searched for the near-infrared (NIR) counterparts of the X-ray sources
in their ESO - NTT or 2MASS observations using the
position measurement from the \emph{Chandra} observations.
The NIR \emph{K$_{\mathrm{s}}$} waveband images of our targets along with their 
\emph{Chandra} images are shown in Figure 
\ref{f3}. The black circles in this figure represent 
the \emph{XMM-Newton} error circles on the position and the white 
circles represent the \emph{Chandra} error circles
except for AX J1820.5--1434 where the \emph{XMM-Newton} error circle 
(see larger one) is also shown in white colour for clarity. 
As can be seen in this figure, we have very likely identified the 
NIR counterparts of AX J1700.1-4157, AX J1740.1-2847 and AX J1749.2-2725
in the ESO - NTT observations. For the remaining two sources - 
AX J1820.5-1434 and AX J1832.3-0840, a bright and a faint NIR counterpart 
respectively is found in the 2MASS observations. The sources identified
in the 2MASS observations could be a combination of a few nearby stars
which might not get resolved due to the poor resolution of the observations
and therefore the probable counterparts of AX J1820.5-1434 and AX J1832.3-0840
are uncertain.
The position and the \emph{J}, \emph{H} and 
\emph{K$_{\mathrm{s}}$} magnitudes of the identified sources 
are listed in Table \ref{t7}. 

\begin{figure*}
\centering
\medskip
\includegraphics[height=7.5cm,width=15.5cm]{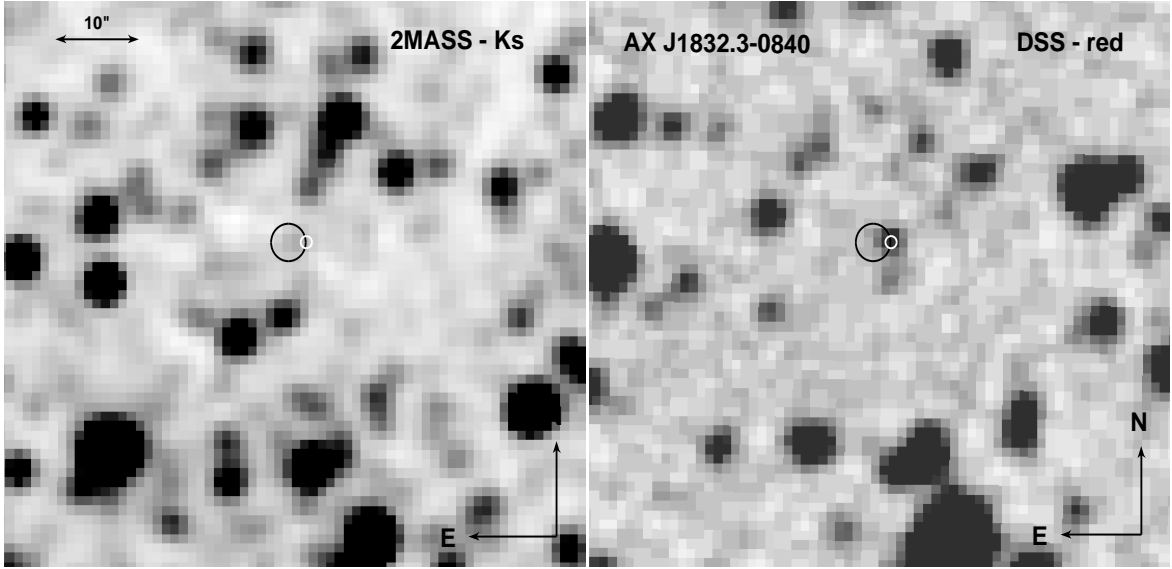}
\caption{\emph{Left} : 2MASS K$_\mathrm{s}$ waveband image of 
AX J1832.3--0840. \emph{Right} : DSS image of the same source 
in the optical R waveband.} 
\label{ff4}
\end{figure*}

We also searched for the optical counterparts of our sources in the 
Digitized Sky Survey (DSS)\footnote[8]{http://stdatu.stsci.edu/cgi-bin/dss\_form} 
observations. For four of our sources, we identified a possible counterpart
in the DSS observations while no optical counterpart is identified 
for AX J1820.5--1434. However, our ESO - NTT observations clearly
showed that
the optical counterparts found for AX J1700.1--4157, AX J1740.1-2847 
and AX J1749.2-2725 are a combination of a few nearby stars, hence we do 
not claim detection of optical counterparts for these three sources.  
Out of the remaining two sources (AX J1820.5--1434 and AX J1832.3--0840) for which we 
did not have the ESO - NTT observations, we could only identify a possible optical 
counterpart for AX J1832.3--0840 in the DSS observations with magnitudes 
R = 19.58 mag and B = 21.13 mag, coincident with its \emph{Chandra} position
(see Figure \ref{ff4}). In the absence of any high-resolution observations
of this source, we consider this faint optical star as its possible counterpart. 

Using the neutral hydrogen column density (N$_{\mathrm{H}}$) measured from 
the absorbed powerlaw model fit to the X-ray spectra of our targets (Table 
\ref{t3}), we calculated the extinction towards them in the \emph{J}, 
\emph{H} and \emph{K$_{\mathrm{s}}$} wavebands following \citet{Predehl1995} 
and \citet{Fitzpatrick1999} and hence calculated the dereddened magnitudes (see Table 
\ref{t8}). Here we assume that the companion star experiences the same 
N$_{\mathrm{H}}$ as observed for the X-ray source (i.e., no local absorption)
which also allows us to put an upper limit on the extinction towards these sources.

On the basis of the extinction-free magnitudes and assuming that the sources
are in our Galaxy (see \citealt{Kaur2009} for the details about this method), 
we tentatively suggest that the 
NIR counterpart of AX J1700.1-4157, AX J1740.1-2847 and AX J1832.3-0840
to be a low-mass star while of AX J1749.2-2725 to be a high mass star. 
The only source for which we did not have the ESO - NTT observations 
and which was not detected in the optical - DSS observations - AX J1820.5--1434, 
showed a bright NIR counterpart in the 2MASS observations. 
Assuming that its counterpart is a single star, we suggest it to be an 
early-type star. 
Even, using the extinction calculated from Galactic N$_{\mathrm{H}}$ \citep{Dickey1990}, 
this source would be compatible with an early-type star.
For AX J1749.2-2725 also, a part of the N$_{\mathrm{H}}$ measured 
could be local to it, however with the Galactic N$_{\mathrm{H}}$, this 
source remains as an early-type star with the given magnitudes. 

We also found a possible tentative relation  
between the pulse period and the measured N$_\mathrm{H}$ of 
these sources. Although we have only 
six sources (including SAX J1324.4--6200 from \citealt{Kaur2009}), 
it seems that these sources tend to form two separate groups -  
having large spin period ($>$ 700 s) and a small N$_\mathrm{H}$ ($<$ 10$^{22}$ 
cm$^{-2}$) and vice-versa, shown in Figure \ref{f4}. 
Our analysis also showed that the sources with large pulse periods have
strong Fe emission line in their spectra (shown  
with `open circles with dot inside' in Figure \ref{t4}) while the other 
sources did not have any signature of it (shown with `filled squares' 
in the same figure).

\begin{table*}
\centering
\caption{Neutral hydrogen column density and the dereddened magnitudes of X-ray pulsators
in  \emph{J}, \emph{H} and \emph{K$_{\mathrm{s}}$} wavebands}
\medskip
\label{t8}
\begin{tabular}{lccccccccc}
\hline
\hline
Object & N$_\mathrm{H}$ & N$^{g}_\mathrm{H}$ &  \emph{J} & \emph{H} & \emph{K$_{\mathrm{s}}$}\\ 
       &  ($\times$ 10$^{22}$ cm$^{-2}$) &  ($\times$ 10$^{22}$ cm$^{-2}$) & mag & mag & mag \\  \hline
AX J1700.1--4157   & 0.5      &  1.8     &  16.26 $\pm$ 0.04  & 16.54 $\pm$ 0.05 & 17.04 $\pm$ 0.13 \\
AX J1740.1--2847   & 1.1      &  0.9     &  14.45 $\pm$ 0.03  & 14.71 $\pm$ 0.07 & 14.86 $\pm$ 0.10 \\
AX J1749.2--2725   & 13.2     &  1.5     &  not detected & 4.25 $\pm$ 0.07 & 6.58 $\pm$ 0.02 \\ 
AX J1820.5--1434   & 8.4      &  1.8     &  2.38 $\pm$ 0.15 & 5.21 $\pm$ 0.07 & 6.29 $\pm$ 0.04 \\
AX J1832.3--0840   & 1.0      &  1.8    &  15.57 $\pm$ 0.23  & 15.27 $\pm$ 0.21 & 15.42 $\pm$ 0.34 \\
\hline
\end{tabular}
\flushleft
NOTE : For AX J1749.2-2725 and AX J1820.5--1434, the measured N$_{\mathrm{H}}$ is much greater than the Galactic N$^{g}_\mathrm{H}$. \\ 
{N$_\mathrm{H}$ - neutral hydrogen column density measured using powerlaw model fit to the
X-ray spectrum,\\
N$^g_\mathrm{H}$  -  neutral hydrogen column density in the direction of the given source
from \citet{Dickey1990} }\\
\end{table*}

\begin{figure*}
\centering
\medskip
\resizebox{10cm}{!}{\includegraphics[angle=-90]{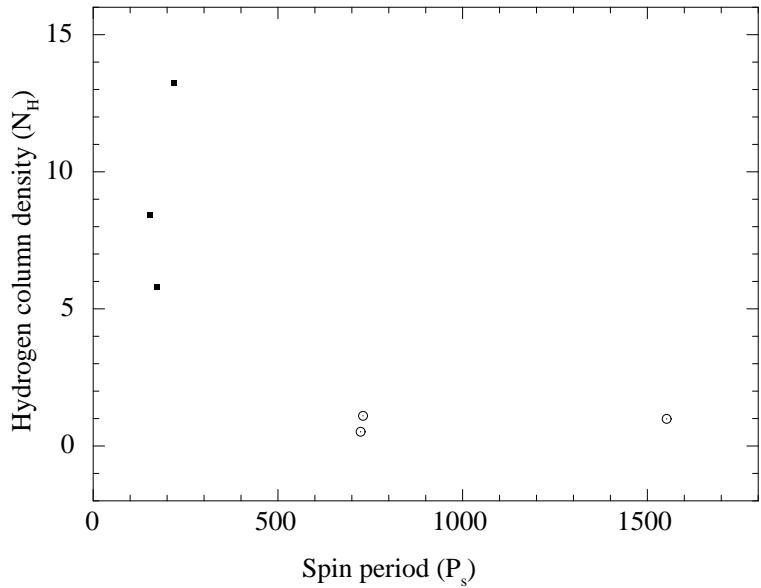}}
\caption{The spin period (P$_\mathrm{s}$) versus the measured neutral hydrogen 
column density (N$_\mathrm{H}$) of our sources. The sources represented by symbol `open
circles with dot inside' have Fe emission lines in their X-ray spectrum while 
the sources represented by symbol `filled square' do not have Fe emission lines.}
\label{f4}
\end{figure*}

\section{Discussion}
\label{discussion}

In this paper, we aimed to find the possible counterparts of five low-luminosity X-ray
pulsators and tentatively identify their  nature. To achieve this goal, we obtained \emph{Chandra} 
observations to measure the position of these sources with a sub-arc second accuracy which helped
us to find their counterparts in the near-infrared/optical observations. We also obtained
\emph{XMM-Newton} observations of these sources to study their X-ray pulsations, 
pulse profiles and spectral parameters. With the help of our \emph{XMM-Newton} observations,
we have determined spin period derivative of AX J1820.5--1434 but only
upper limits for the other sources. Our sources are likely neutron star binaries (with a high mass or a
low mass companion) or accreting white dwarfs.

Most of the known slow X-ray pulsars are found among HMXBs except a very few which
have low-mass companions. The known slow (P$_\mathrm{s}$ $>$ 0.1 s) X-ray pulsars with low-mass 
companions are Her X-1, 4U 1626--67, 
GRO 1744-28, 4U 1822-37 and GX 1+4  (\citealt{Bildsten1997}, \citealt{Jonker2001}). 
Four of these LMXB pulsars have spin periods between 
0.5 - 10 s while only GX 1+4 has a spin-period of 140 s but also a red giant star as its companion. 
However, the spin period of X-ray pulsars containing high mass companions are 
observed upto thousands of seconds \citep{Bildsten1997}. 
The X-ray pulsars (both in LMXBs and HMXBs) usually display a hard X-ray spectrum 
well fitted with a powerlaw index, $\Gamma$ $\sim$ 1.0, or with a blackbody model 
of temperature, $kT$ $\sim$ 2 keV. Sometimes, their X-ray spectrum is also characterized 
by a strong neutral Fe emission line at 6.4 keV and very rarely ionized Fe lines at 6.7 
and 6.9 keV \citep{Ebisawa1996}. These sources show both spin-up and spin-down
behaviour on time-scales from a few days to a few years \citep{Bildsten1997}.
The fastest spin-period derivative observed among neutron star binary pulsars
is $\sim$ 10$^{-7}$ s s$^{-1}$ for GX 1+4 \citep{Elsner1985}.
On the other hand, intermediate polars (IPs; a sub-class of white dwarf binary systems; 
\citealt{Warner2003}) have been observed with spin periods in the
range of a few hundred seconds to a few thousand seconds \citep{Kuulkers_book}.
The observed X-ray spectra of most of these systems are well fitted with
a thin plasma or bremsstrahlung model with temperatures of $\sim$ 1 - 30 keV 
and characterized by strong Fe emission lines
at 6.4, 6.7 and 6.9 keV \citep{Ezuka1999}. In most of them, the X-ray 
spectrum also fits well with powerlaw index, $\Gamma$ $\le$ 1.0 \citep{Hong2009}. 
Most of the white dwarfs in IPs spin-up with a 
typical spin-period derivative of 10$^{-11}$ s s$^{-1}$ except a few of them
which spin-down. The largest spin-period derivative observed in these 
systems is 1.1 $\times$  10$^{-10}$ s s$^{-1}$ \citep{Mason1997}.

{\flushleft{\bf{AX J1700.1--4157}}}

AX J1700.1--4157, which has a pulse period of 714 s, was discovered 
from \emph{ASCA} observations performed on September 16, 1997 with a
hard X-ray spectrum ($\Gamma$ $\sim$ 0.7; \citealt{Torii1999}). 
It is unlikely a LMXB pulsar due to its 
large pulse period. 
The timing and spectral parameters measured from our observations 
are consistent with the previous observations reported by \citet{Torii1999}.
However, we detected a strong Fe emission line at 6.7 keV with an equivalent 
width of 580 eV during our observations, for which an upper limit of 
1150 eV was given by \citet{Torii1999}. 
From our NIR observations, we estimate a low mass star to be 
the counterpart of this source which in combination with the detection 
of Fe emission line would favor an IP nature of this source.

{\flushleft{\bf{AX J1740.1--2847}}}

AX J1740.1--2847 was discovered with a pulse period of 729 $\pm$ 14 s
from observations performed on September 7 - 8, 1998 using \emph{ASCA}
and with a hard X-ray spectrum ($\Gamma$ $\sim$ 0.7; \citealt{Sakano2000}). 
The large pulse period of this source makes it  
unlikely to be a LMXB pulsar. The timing and spectral parameters measured from our observations
are consistent with the previous measurement except that we detected a 
strong Fe 6.4 keV line in the X-ray spectrum with an equivalent width 
of 998 eV which was not detected during the  \emph{ASCA} observations
and an upper limit of 500 eV was given on its equivalent width \citep{Sakano2000}.
With the given NIR magnitudes,  we suggest a 
low mass star to be the possible counterpart of this source, favoring 
an IP interpretation. 

{\flushleft{\bf{AX J1749.2--2725}}}

AX J1749.2--2725, discovered with a pulse period of 220.38 $\pm$ 0.20 s
during the \emph{ASCA} observations performed on March 26, 1995, 
was detected with the X-ray spectrum having powerlaw index of 1.0 \citep{Torii1998}. 
During our observations, AX J1749.2--2725 was detected 
with similar timing and spectral parameters as were reported by \citet{Torii1998}.
No Fe emission line was detected in the X-ray spectrum during the previous as well as 
the present observations. 
From our NIR observations, we infer a high mass star as a possible counterpart
which favors a HMXB pulsar nature for this source.

{\flushleft{\bf{AX J1820.5--1434}}}

AX J1820.5--1434 was detected with a pulse period of 152.26 $\pm$ 0.04 s
from the \emph{ASCA} observations made on April 8 - 12, 1997
and with a hard X-ray spectrum ($\Gamma$ $\sim$ 0.9; \citealt{Kinugasa1998}). 
The observed timing
and spectral parameters from our observations are consistent 
with those reported by \citet{Kinugasa1998}. The accurate spin-period 
measurement from our observations, in combination with the 
previous spin-period measurement (Table \ref{tt1}) resulted in a 
spin-period derivative determination of this source  
of (3.00 $\pm$ 0.14) $\times$ 10$^{-9}$ s s$^{-1}$.
The largest spin-period derivative measured in IPs 
is 1.1 $\times$ 10$^{-10}$ s s$^{-1}$ \citep{Mason1997}. Thus, this source
is unlikely a white dwarf system. 
Although we identified a bright NIR counterpart of this source in the 
2MASS observations, we caution against drawing any
conclusions on the basis of it since it could be a combination
of a few nearby stars overlapped. The high spin-down rate of this 
source favors a neutron star nature, but we cannot distinguish 
between a low mass or a high mass X-ray binary.

{\flushleft{\bf{AX J1832.3--0840}}}

AX J1832.3--0840 which has a pulse period of 1550 s, was discovered 
during observations performed on October 11, 1997 using \emph{ASCA},
with a hard X-ray spectrum ($\Gamma$ $\sim$ 0.8; \citealt{Sugizaki2000}).
During these observations, a strong 6.7 keV Fe emission line 
was detected in the X-ray spectrum and the spectrum 
was also satisfactorily fitted with a thermal-equilibrium plasma model \citep{Raymond1977}.
During our observations, we measured the similar timing
and the spectral parameters with respect to the previous observations. However,
we detected three Fe emission lines in the X-ray spectrum of this source 
(see Figure \ref{f2}), which are typically seen in IPs. Our X-ray spectrum 
satisfactorily fitted with a bremsstrahlung model of temperature $\sim$ 30 keV, which 
is common among IPs and is rare among X-ray pulsars. We identified a
faint counterpart of this source in the DSS as well as  2MASS observations,
which indicate a low-mass star as its counterpart. Therefore, we suggest 
this source to be likely an IP.

\section{Conclusions}
\label{conclusions}

Our multiwavelength observations have helped us to identify the very likely 
near-infrared/optical 
counterparts for most of our targets. The X-ray flux
of all the sources is found consistent with their previous measurements 
and the non-detection of any period of quiescence in them  
confirmed their persistent nature. 
The pulse profiles and pulse fractional amplitudes from our observations 
are found to be consistent with the previous observations, 
indicating that these are stable systems. 
Our observations also suggest that these sources might form two different groups 
-  with large pulse period having small N$_H$ and detection of Fe lines
and small pulse period  having large N$_H$ and non-detection of any Fe emission
line (Figure \ref{f4}). With the help of our multiwavelength observations, we suggest
AX J1749.2--2725 and AX J1820.5--1434 to be accreting neutron star systems 
while the remaining three sources could be IPs. The NIR spectroscopic observations 
of these sources would help us to find the exact nature of the counterparts 
and to unveil their true nature.


\begin{thebibliography}{32}
\expandafter\ifx\csname natexlab\endcsname\relax\def\natexlab#1{#1}\fi
\expandafter\ifx\csname url\endcsname\relax
  \def\url#1{{\tt #1}}\fi
\expandafter\ifx\csname urlprefix\endcsname\relax\def\urlprefix{URL }\fi

\bibitem[{Aldcroft} et~al.(2000)]{Aldcroft2000}
{Aldcroft} T.L., {Karovska} M., {Cresitello-Dittmar} M.L., {Cameron} R.A. and
  {Markevitch} M.L.
\newblock 2000, in {Truemper} J.E. and {Aschenbach} B., eds, Proc. SPIE Vol. 4012, 
\emph{X-ray Optics, Instruments, and Missions III}. SPIE, Bellingham, p. 650

\bibitem[{Bildsten} et~al.(1997)]{Bildsten1997}
{Bildsten} L., {Chakrabarty} D., {Chiu} J. et~al.
\newblock 1997, \apjs, 113, 367 

\bibitem[{Dickey} and {Lockman}(1990)]{Dickey1990}
{Dickey} J.M. and {Lockman} F.J.
\newblock 1990, \araa, 28, 215 

\bibitem[{Ebisawa} et~al.(1996)]{Ebisawa1996}
{Ebisawa} K., {Day} C.S.R., {Kallman} T.R. et~al.
\newblock 1996, \pasj, 48, 425 

\bibitem[{Elsner} et~al.(1985)]{Elsner1985}
{Elsner} R.F., {Weisskopf} M.C., {Apparao} K.M.V. et~al.
\newblock 1985, \apj, 297, 288 

\bibitem[{Ezuka} and {Ishida}(1999)]{Ezuka1999}
{Ezuka} H. and {Ishida} M.
\newblock 1999, \apjs, 120, 277 

\bibitem[{Fitzpatrick}(1999)]{Fitzpatrick1999}
{Fitzpatrick} E.L.
\newblock 1999, \pasp, 111, 63 

\bibitem[{Hong} et~al.(2009)]{Hong2009}
{Hong} J.S., {van den Berg} M., {Laycock} S., {Grindlay} J.E. and {Zhao} P.
\newblock 2009, \apj, 699, 1053 

\bibitem[{Hulleman} et~al.(1998)]{Hulleman1998}
{Hulleman} F., {in 't Zand} J.J.M. and {Heise} J.
\newblock 1998, \aap, 337, L25 

\bibitem[{Jonker} and {van der Klis}(2001)]{Jonker2001}
{Jonker} P.G. and {van der Klis} M.
\newblock 2001, \apj, 553, L43

\bibitem[{Kaur} et~al.(2009)]{Kaur2009}
{Kaur} R., {Wijnands} R., {Patruno} A. et~al.
\newblock 2009, \mnras, 394, 1597

\bibitem[{Kinugasa} et~al.(1998)]{Kinugasa1998}
{Kinugasa} K., {Torii} K., {Hashimoto} Y. et~al.
\newblock 1998, \apj, 495, 435

\bibitem[{Kuulkers} et~al.(2006)]{Kuulkers_book}
{Kuulkers} E., {Norton} A., {Schwope} A. and {Warner} B.
\newblock 2006, in Compact stellar X-ray sources, ed. W. H. G.
Lewin \& M. van der Klis (Cambridge: Cambridge Univ. Press), 421 

\bibitem[{Lin} et~al.(2002)]{Lin2002}
{Lin} X.B., {Church} M.J., {Nagase} F. and {Ba{\l}uci{\'n}ska-Church} M.
\newblock 2002, \mnras, 337, 1245 

\bibitem[{Makishima}(1986)]{Makishima1986}
{Makishima} K.
\newblock in Mason, K. O., Watson, M. G., \& White, N. E. eds, \emph{The Physics of Accretion onto Compact Objects} Springer-Verlag, Berlin, p. 249

\bibitem[{Mason}(1997)]{Mason1997}
{Mason} K.O.
\newblock 1997, \mnras, 285, 493 

\bibitem[{Mewe} et~al.(1995)]{Mewe1995}
{Mewe} R., {Kaastra} J.S., {Schrijver} C.J., {van den Oord} G.H.J. and
  {Alkemade} F.J.M.
\newblock 1995, \aap, 296, 477

\bibitem[{Misaki} et~al.(1996)]{Misaki1996}
{Misaki} K., {Terashima} Y., {Kamata} Y. et~al.
\newblock 1996, \apjl, 470, L53 

\bibitem[{Okazaki} and {Negueruela}(2001)]{Okazaki2001}
{Okazaki} A.T. and {Negueruela} I.
\newblock 2001, \aap, 377, 161 

\bibitem[{Pfahl} et~al.(2002)]{Pfahl2002}
{Pfahl} E., {Rappaport} S., {Podsiadlowski} P. and {Spruit} H.
\newblock 2002, \apj, 574, 364 

\bibitem[{Predehl} and {Schmitt}(1995)]{Predehl1995}
{Predehl} P. and {Schmitt} J.H.M.M.
\newblock 1995, \aap, 293, 889 

\bibitem[{Raymond} and {Smith}(1977)]{Raymond1977}
{Raymond} J.C. and {Smith} B.W.
\newblock 1977, \apjs, 35, 419 

\bibitem[{Reig} and {Roche}(1999)]{Reig1999}
{Reig} P. and {Roche} P.
\newblock 1999, \mnras, 306, 100 

\bibitem[{Sakano} et~al.(2000)]{Sakano2000}
{Sakano} M., {Torii} K., {Koyama} K., {Maeda} Y. and {Yamauchi} S.
\newblock 2000, \pasj, 52, 1141 

\bibitem[{Str{\"u}der} et~al.(2001)]{Struder2001}
{Str{\"u}der} L., {Briel} U., {Dennerl} K. et~al.
\newblock 2001, \aap, 365, L18 

\bibitem[{Sugizaki} et~al.(2000)]{Sugizaki2000}
{Sugizaki} M., {Kinugasa} K., {Matsuzaki} K. et~al.
\newblock 2000, \apjl, 534, L181 

\bibitem[{Sugizaki} et~al.(2001)]{Sugizaki2001}
{Sugizaki} M., {Mitsuda} K., {Kaneda} H. et~al.
\newblock 2001, \apjs, 134, 77

\bibitem[{Torii} et~al.(1998)]{Torii1998}
{Torii} K., {Kinugasa} K., {Katayama} K. et~al.
\newblock 1998, \apj, 508, 854 

\bibitem[{Torii} et~al.(1999)]{Torii1999}
{Torii} K., {Sugizaki} M., {Kohmura} T., {Endo} T. and {Nagase} F.
\newblock 1999, \apjl, 523, L65 

\bibitem[{Turner} et~al.(2001)]{Turner2001}
{Turner} M.J.L., {Abbey} A., {Arnaud} M. et~al.
\newblock 2001, \aap, 365, L27 

\bibitem[{van den Berg} et~al.(2004)]{vandenberg2004}
{van den Berg} M., {Tagliaferri} G., {Belloni} T. and {Verbunt} F.
\newblock 2004, \aap, 418, 509

\bibitem[{Warner}(2003)]{Warner2003}
{Warner} B.
\newblock 2003, \emph{Cataclysmic Variable Stars}. Cambridge Univ. Press, Cambridge

\end{thebibliography}
\end{document}